\documentclass[11pt]{article}

\usepackage{amssymb,amsmath}  

\usepackage[square]{natbib}

\begin{document}  
\begin{center}  
\large\textbf{Gauge Symmetry Breaking in Gauge Theories---In Search of Clarification}\vspace{0.7cm}\\  

\normalsize  
\textbf{Simon Friederich}\vspace{0.2cm}\\  

\texttt{friederich@uni-wuppertal.de}\vspace{0.2cm}\\  

Universit\"at Wuppertal, Fachbereich C -- Mathematik und Naturwissenschaften, Gau\ss str. 20, D-42119 Wuppertal, Germany\vspace{0.2cm}\\  

\end{center}  
\noindent\textbf{Abstract:}  
The paper investigates the spontaneous breaking of gauge symmetries in gauge theories from a philosophical angle, taking into account the fact that the notion of a spontaneously broken local gauge symmetry, though widely employed in textbook expositions of the Higgs mechanism, is not supported by our leading theoretical frameworks of gauge quantum theories. In the context of lattice gauge theory, the statement that local gauge symmetry cannot be spontaneously broken can even be made rigorous in the form of \textit{Elitzur's theorem}. Nevertheless, gauge symmetry breaking does occur in gauge quantum field theories in the form of the breaking of remnant subgroups of the original local gauge group under which the theories typically remain invariant after gauge fixing. The paper discusses the relation between these instances of symmetry breaking and phase transitions and draws some more general conclusions for the philosophical interpretation of gauge symmetries and their breaking. 
\vspace{0.5cm}\\  

\section{Introduction}  
The interpretation of symmetries and symmetry breaking has been recognized as a central topic in the philosophy of science in recent years. Gauge symmetries, in particular, have attracted a considerable amount of interest due to the central role they play in our most successful theories of the fundamental constituents of nature. The standard model of elementary particle physics, for instance, is formulated in terms of gauge symmetry, and so are its most discussed extensions. By most accounts, gauge symmetries are not empirical but purely formal features of gauge theories\footnote{See \citep{healey}, Chapter 6, for an in-depth philosophical exposition of the distinction between empirical and purely formal symmetries that defends the standard account of gauge symmetries as purely formal.} in the sense that different configurations of the fields involved represent identical physical situations if they are related by gauge symmetry. Saying that gauge symmetries are purely formal in that sense is not equivalent to denying that they may have important empirical significance in a more indirect way. For recent philosophical work on the interpretation of gauge symmetries see, for instance, \citep{redhead}, \citep{brading}, \citep{lyreU1}, \citep{healey}, \citep{greaves} (the latter openly critical of the standard account of gauge symmetries as having no direct empirical significance, which is also adopted here).  

The present paper focuses on a particular aspect of gauge symmetries, namely, the notion of a spontaneously broken gauge symmetry. This is a notion that may seem puzzling at first glance, for it seems natural to ask what it might mean to spontaneously break a purely formal symmetry that exists only on the level of our description of physical reality, not on the level of physical reality itself. The notion of a spontaneously broken gauge symmetry is not an exotic notion, however, for it is widely regarded as playing a crucial role in the generation of particle masses in the standard model of particle physics by the Higgs mechanism. Although it is almost universally accepted, the received view of the Higgs mechanism as a case of broken local gauge symmetry has been criticized by both physicists and philosophers of physics, see \citep{thooft}, \citep{earmanb}, \citep{healey}, \citep{lyre}. 't Hooft, for instance, criticizing it from the point of view of a physicist, claims that the notion of a spontaneously broken local gauge symmetry is ``something of a misnomer''\footnote{See \citep{thooft}, p.\ 697.}, while Earman, from the point of view of a philosopher, expresses qualms concerning the Higgs mechanism as a spontaneously broken local gauge symmetry on grounds that ``a genuine physical property like mass cannot be gained by eating descriptive fluff, which is just what gauge is.''\footnote{See \citep{earmanb}, p.\ 1239.} Worries like these about the standard picture of the Higgs mechanism as a spontaneously broken local gauge symmetry are aggravated by a result of lattice gauge theory known as \textit{Elitzur's theorem} (see Section 5), according to which local (gauge) symmetry cannot be spontaneously broken at all.  

In order to develop an adequate perspective on the status of spontaneous symmetry breaking in quantized gauge theories, Earman proposes that the question be tackled by means of reduced phase space quantization, an approach in which, in contrast to the standard approach discussed in Sections 5 and 6 of this paper, gauge orbits (that is, gauge field configurations related by gauge symmetry) are quotiented out before the resulting unconstrained system of variables is subjected to a quantization procedure (see \citep{earmanb}). Independently of the success of this enterprise\footnote{See \citep{struyveold}, Section 8 and, in particular, \citep{struyve} for recent progress on it.}, it seems reasonable to ask whether the puzzles surrounding the notion of spontaneous symmetry breaking in gauge theories might not be resolvable within the standard ``Lagrangean'' framework of quantum field theories, as it is actually used by those working in the field of high energy physics. My aim in the present paper will be to show that this can indeed be done. A proper assessment of the role of symmetry breaking in gauge theories that does not merely recite the standard narrative of the Higgs mechanism as a spontaneously broken local gauge symmetry, arguably, can be given on the basis of the conventional approach to quantum field theory alone. 

The rest of this paper is organized as follows: Section 2 recalls some basic features of the concepts of (gauge) symmetry and (gauge) symmetry breaking, and Section 3 discusses the characterization of symmetry breaking as a ``natural phenomenon'' proposed by Liu and Emch\footnote{See \citep{liu}, p.\ 153.} and considers in which sense it applies to cases where the broken symmetry is a \textit{gauge} symmetry. Sections 4 and 5 assess the fate of the notion of \textit{local} symmetry breaking in gauge theories, which, as argued in Section 4, makes sense at the classical level but is not supported by present-day quantized gauge theories, as discussed in Section 5. Section 6 discusses the breaking of post-gauge fixing remnant \textit{global} gauge symmetries and their relation to transitions between distinct physical phases. It is argued that there seems to be no fixed connection between these instances of symmetry breaking and phase transitions in that the distinction between broken and unbroken symmetries does not in general line up with a distinction between distinct physical phases. Section 7 turns to the more general philosophical relevance of these findings by considering their implications for claims brought forward in the literature on philosophical aspects of gauge symmetries and their breaking. The paper closes in Section 8 with a brief concluding remark.

\section{Symmetries, gauge symmetries, and symmetry breaking}  
In this section, I give a brief review of the concepts in terms of which the questions discussed in this paper are formulated. The concepts are those of symmetry, gauge symmetry, symmetry breaking, and gauge symmetry breaking. Readers who are familiar with these notions can skip the section without loss, perhaps apart from the last three paragraphs, which review the phenomenon of Bose-Einstein condensation in a free Bose gas in terms of broken gauge symmetry.  

A \textit{symmetry} $\alpha$ of a classical system is a transformation $\alpha:\gamma\mapsto\alpha\gamma$ of the (coordinate) variables in terms of which its configurations $s_\gamma$ are individuated that induces an automorphism $\alpha:s_\gamma\mapsto\alpha s_\gamma\equiv s_{\alpha\gamma}$ which commutes with its time evolution. For a quantum system, a symmetry is an automorphism of the observables or canonical variables which preserves all algebraic relations. In analogy to the classical case, possible states of the system are individuated in terms of the expectation values they ascribe to these quantities. Time evolution, in the Heisenberg picture, counts as an algebraic relation among others, so the invariance of all algebraic relations under a symmetry in the Heisenberg picture implies that the symmetry commutes with the dynamics of the system.  

\textit{Gauge theories} are defined in terms of an action $S$ that is invariant under transformations which typically correspond to the elements of an infinite dimensional Lie group and depend on a finite number of arbitrary functions. As follows from Noether's second theorem (see \citep{noether}), the equations of motion apparently fail to be deterministic in this case in that their solutions involve arbitrary functions of space-time. This means, in particular, that any configuration of the coordinate variables at a given initial time $t_0$ does not uniquely determine the configuration of variables at a later time $t_1$. In the \textit{physical} interpretation of gauge theories, however, determinism can be restored by assuming that variable configurations that are related by the symmetry represent identical physical situations. The symmetry is referred to as a \textit{gauge symmetry} in this case. Classical electromagnetism is a paradigm example of a gauge theory in that (assuming the relativistic formulation in terms of four-vector fields) its action is invariant under local gauge transformations of the four-vector potential $A_\mu(x)$ having the form  
\begin{eqnarray}\label{localgauge}  
A_\mu(x)\mapsto A_\mu(x)-\frac{1}{e}\partial_\mu\alpha(x)\,.  
\end{eqnarray}  
Only functions of the gauge fields that are invariant under gauge transformations of the form (\ref{localgauge}) correspond to physical quantities. The inertial frame-dependent electric and magnetic fields $\mathbf E(x)$ and $\mathbf B(x)$, obtained from $A_\mu(x)$ by taking certain derivatives, are examples of such quantities, and only these, not the gauge fields themselves nor any other gauge-dependent quantities, are observable. The quantization of classical gauge theories is most conveniently carried out by means of functional integral quantization, which provides the basis for the vast majority of physicists' studies of quantum gauge theories. The framework of functional integral quantization for gauge theories is briefly sketched in Section 5 and used as the background of the discussion of gauge symmetry breaking in the quantum context in Sections 5 and 6. 

Symmetries that are gauge symmetries in the sense just discussed are often referred to as \textit{local} symmetries, alluding to the fact that the parameters of symmetry transformations can be chosen ``locally'', that is, independently of each other for distinct space-time regions (see, for instance, the freedom in the choice of $\alpha(x)$ in Eq.\ (\ref{localgauge})). However, the idea that variable configurations related by symmetry correspond to identical physical situations applies also in contexts where the symmetry transformations depend on only finitely many parameters (though there may be disputes on which contexts these are), and one speaks of ``global gauge symmetries'' with regard to these cases, in contrast to the ``local'' gauge symmetries discussed before. Only theories that are formulated in terms of \textit{local} gauge symmetry are commonly referred to as gauge theories, however. The present paper adopts this standard use of terminology, understanding ``gauge symmetry'' to refer to both local and global symmetries having the property that variable configurations related by symmetry represent identical physical situations and ``gauge theory'' to refer to theories formulated in terms of \textit{local} gauge symmetry only.

Having reviewed the notions of symmetry in general and of gauge symmetry in particular, I now turn to the notion of \textit{spontaneous symmetry breaking} (``SSB'' in what follows). The basic idea behind this concept is that the mapping of the state space onto itself which is induced by a symmetry transformation does not in general map each single state onto itself. Put differently, this means that the state of a physical system need not have all the symmetries which the laws of motions governing its behaviour have. States for which this is the case are candidates for exhibiting SSB, and for the purposes of the present paper, where the focus is on ground states and thermal states of theories with infinitely many degrees of freedom, one may simply identify them with the spontaneously symmetry breaking ones.\footnote{See \citep{strocchi} for a rigorous textbook account of SSB that avoids both unnecessary technicalities and misleading simplifications. Roughly speaking, for a state to exhibit SSB in the rigorous sense specified in that work, it needs to take an infinite amount of energy to transform the system from one asymmetric state into another. This is the reason why realistic systems (that is, systems without any potential barriers of infinite height) need to have infinitely many degrees of freedom to exhibit SSB. Furthermore, it does not suffice for a non-symmetric state to differ only \textit{slightly} from a symmetric one (in the sense in which, say, a single particle state differs from a zero-particle, fully symmetric, vacuum state) to qualify as symmetry breaking. Cases like these are automatically excluded by the criterion in terms of the algebraic approach to quantum theories stated in the next paragraph.}

For quantum theories, the basic idea behind the concept of SSB just sketched can be turned into a precise criterion using the language of the algebraic approach to quantum theories. One defines that for a symmetry $\alpha$ of the algebra of observables of a system to be spontaneously broken by a state $\omega$, the GNS representations associated with the states $\omega$ and $\alpha^*\omega$ (defined by $\alpha^*\omega(A)=\omega(\alpha(A))$) have to be unitarily inequivalent.\footnote{For accessible introductions to the notions of algebras of observables, their representations, the unitary (in-) equivalence of representations, and a state's GNS representation, see, for instance, \citep{ruetsche}, Chapter 13, and \citep{strocchi}.} Intuitively, this means that the states $\omega$ and $\alpha^*\omega$ cannot be written in the form of density matrices in one and the same Hilbert space $\mathcal H$. An expectation value $\omega(A)$ of an observable $A$ for which  
\begin{eqnarray}  \label{orderparam}
\omega(A)\neq\alpha^*\omega(A)  
\end{eqnarray}  
is called a symmetry breaking order parameter for the symmetry $\alpha$ in the state $\omega$. Situations where the symmetry $\alpha$ is broken are characterized by the fact that this quantity is nonzero, whereas it vanishes for states that are symmetric with respect to $\alpha$. Symmetry breaking order parameters in the sense of Eq.\ (\ref{orderparam}) can be used to \textit{define} SSB in contexts where the algebraic criterion does not apply in that the quantum theory at issue is not formulated in terms of the algebraic approach. This holds, for instance, for the application of the concept of SSB in the framework of the functional integral formulation of quantum theories in which the quantization of gauge theories is most commonly carried out (see Section 5 of this paper for more details).

The notion of a spontaneously broken \textit{gauge} symmetry may seem puzzling at first sight. As formulated by Smeenk, ``[i]f gauge symmetry merely indicates descriptive redundancy in the mathematical formalism, it is not clear how spontaneously breaking a gauge symmetry could have any physical consequences, desirable or not.''\footnote{See \citep{smeenk}, p.\ 488. See \citep{earmanb}, Section 9, for a similar way of putting the challenge.} Part of the aim of the present paper is to remove the puzzlement expressed in Smeenk's statement and to elucidate the physical significance of SSB for gauge symmetries. For the purposes of the present section it suffices to clarify the notion of a spontaneously broken gauge symmetry at a technical level, and to do so, the account just given for SSB on the level of observables must be generalized by extending the algebra of observables to an algebra of canonical variables that are not themselves physical observables. A simple example of a quantum theory with a spontaneously broken gauge symmetry is that of Bose-Einstein condensation of a non-relativistic free Bose gas in the thermodynamic limit at zero temperature.\footnote{The following presentation relies on \citep{strocchi}, Chapter 7.2. See also Chapters 13.3 and 13.4 of \citep{strocchi} for further details.} Since this theory is formulated in terms of \textit{global}, rather than local, gauge symmetry, it does not qualify as a gauge theory, but the spontaneous breaking of a gauge symmetry can nevertheless nicely be illustrated with it. In this case, the canonical variables are (quantum) fields $\psi(x)$, and the system has infinitely many pure ground states $\Omega_\theta$, labelled by different values of an angle variable $\theta$, all of which assign a nonzero expectation value to the (improper) field operator $\psi(x)$:  
\begin{eqnarray}  \label{Omega}
\Omega_\theta(\psi)=\sqrt{n}e^{i\theta},\hspace{1cm}\theta\in[0,2\pi)\,, 
\end{eqnarray}
where $n$ is the average density $n=|\Omega_\theta(\psi)|^2$.

Physically, all states $\Omega_\theta$ defined in Eq.\ (\ref{Omega}) are equivalent to each other in that they (and their mixtures) yield the same expectation values for all observable quantities. Gauge symmetry comes into play in the form of an invariance of the dynamics under global gauge transformations of the form  
\begin{eqnarray}\label{globalgauge}  
\psi(x)&\mapsto&\alpha^\lambda(\psi(x))=e^{i\lambda}\psi(x)\,,\nonumber\\  
\psi^*(x)&\mapsto&\alpha^\lambda(\psi^*(x))=e^{-i\lambda}\psi^*(x)\,,
\end{eqnarray}  
where $\lambda\in[0,2\pi)$.  

The states $\Omega_\theta$ are not invariant under these transformations in that  
\begin{eqnarray}  
\Omega_\theta(\alpha^\lambda(\psi))\,  
=\Omega_{\theta+\lambda}(\psi)\,  
\neq\Omega_\theta(\psi)\,  
\end{eqnarray}  
for $\lambda\neq0$, so they break the gauge symmetry and $\Omega_\theta(\psi)$ qualifies as a symmetry breaking order parameter. As an order parameter testifying to the breaking of a \textit{gauge} symmetry it is not itself an observable, though its modulus $\sqrt{n}=|\Omega_\theta(\psi)|$ is. Gauge symmetry breaking is an unavoidable feature of one's description if one wants to describe the free Bose gas at zero temperature in the thermodynamic limit in terms of gauge variables by means of a pure state, but the states $\Omega_\theta$, among which one can choose, are all \textit{physically} equivalent in that they assign the same expectation values to all observables.\footnote{\citep{leggett} provides an illuminating discussion of Bose-Einstein condensation in the absence of the thermodynamic limit that does not operate with the notion of a spontaneously broken gauge symmetry. The assumptions underlying Leggett's approach are more realistic than those of the discussion made here, since the number of atoms in physically realized examples of BEC is not exceedingly large (between roughly $10^{3}$ and $10^{5}$) and there are important inter-particle interactions in these systems. Leggett's ``order parameter'' (see \citep{leggett} Eq.\ (2.2.1)), in terms of which he defines Bose-Einstein condensation, is not a symmetry breaking order parameter in the sense of Eq.\ (\ref{orderparam}).}

\section{Symmetry breaking as a natural phenomenon}  
Spontaneous symmetry breaking, as explained in the previous section, is a feature of states that do not have the all the symmetries of the underlying laws of motions (in theories with infinitely many degrees of freedom). In order to \textit{interpret} the notion thereby defined, let us first focus on cases where the broken symmetry is one of the algebra of observables (that is, not a gauge symmetry), so that its breaking manifests itself as an asymmetry on the level of observables. Having in mind these cases of SSB, Liu and Emch characterize symmetry breaking by means of the non-technical and intuitive notion of a ``natural phenomenon''\footnote{See \citep{liu}, p.\ 153. Liu and Emch focus on \textit{quantum} spontaneous symmetry breaking, specifically, but the characterization of symmetry breaking as a natural phenomenon does not seem to be based on any particular quantum (as opposed to classical) aspects.}, contrasting it with ``merely theoretical concepts'' such as ``renormalization, first- or second- quantization.''\footnote{See \citep{liu}, p.\ 153, fn.\ 14.} Whenever the state of a system as specified in terms of the expectation values of its observables spontaneously breaks some symmetry of the underlying laws of motion, the discrepancy between the symmetries of the state and those of the laws is an objective feature of the physical situation described by that state and not merely an artefact of our description. Liu's and Emch's characterization of SSB as a ``natural phenomenon'' therefore seems adequate for cases of SSB on the level of observables in that the breaking of these symmetries, whenever it occurs, is an objective matter and not merely a conventional or otherwise arbitrary feature of how we represent the physical situation.\footnote{Note that to accept the characterization of SSB on the level of observables in quantum theories as a natural phenomenon, it does not seem necessary to endorse the standard \textit{ontic} view of quantum states as states quantum systems ``are in''. The main reason for adopting the alternative, epistemic, conception of quantum states is that it elegantly avoids the paradoxes of measurement and nonlocality. (See \citep{friederich} for more details and an exploration of how the view might be spelled out in detail.) According to the epistemic conception of quantum states, quantum states reflect the state assigning agents' epistemic relations to these systems, so no such thing as the ``true'' quantum state of a quantum system is acknowledged, and SSB cannot be characterized in terms of quantum systems' ``being in'' quantum states that break some symmetry of the algebra of observables. Nevertheless, proponents of the epistemic conception of states can hold that SSB is a natural phenomenon in that an observable called a ``witness'' of a symmetry of the observables may have a value that, if known, requires the assignment of a state that breaks that symmetry. (For an explanation of the notion of an observable being a ``witness'' for SSB, see \citep{liu}, p.\ 145.) The characterization of SSB in quantum theories as a natural phenomenon seems therefore independent of the question of whether quantum states are conceived of as ontic or non-ontic.}

While SSB on the level of observables seems adequately characterized as a ``natural phenomenon'' in the sense just discussed, the status of SSB on the level of gauge variables seems less clear. The reason for this is that gauge symmetries, as explained in the first section, are purely formal symmetries that have no physical instantiations. Whenever we describe some physical situation in terms of broken gauge symmetry, there is thus no discrepancy between the \textit{physical} symmetries of the situation and those of the underlying laws of motion. This can nicely be seen, for instance, in the case of Bose-Einstein condensation mentioned at the end of the previous section, where the gauge symmetry is broken by any of the states $\Omega_\theta$, but the physical properties of the system, i.\ e., the expectation values of observables, are exactly the same for all $\Omega_\theta$. There is in this case no asymmetry in the physical, gauge invariant, properties of the system which the underlying laws of motion do not have. In just the same sense in which gauge symmetries contrast with empirical symmetries in that they have no physical instantiations gauge symmetry \textit{breaking} seems to be rather an aspect of how we describe a physical situation than an objective feature of the situation itself.

One may feel, however, that to conclude from these considerations that gauge symmetry breaking does not deserve to be characterized as a ``natural phenomenon'' in any reasonable sense would be too rash. More specifically, one may feel that whether some physical system is described in terms of \textit{broken} or \textit{unbroken} gauge symmetry relates directly to objective features of that system. Even though SSB does not seem to be an \textit{intrinsic physical} feature of systems described in terms of broken gauge symmetry in the same way as it is for systems described in terms of a broken symmetry on the level of observables, it might nevertheless be regarded as an \textit{extrinsic physical} feature of these systems in the sense that their physical characteristics may strongly differ from those of systems described in terms of \textit{unbroken} gauge symmetry. Instances of gauge symmetry breaking, one might want to say, deserve to be called ``natural phenomena'' if and only if situations described in terms of broken gauge symmetry are qualitatively different from those described in terms of unbroken gauge symmetry. However, since both the notion of a natural phenomenon and that of a ``qualitative difference'' between physical situations are only intuitive notions, this idea is in need of further qualification and should be made more precise.

A natural way of doing so is to say that gauge symmetry breaking qualifies as a ``natural phenomenon'' just in case the distinction between broken and unbroken gauge symmetry lines up completely with a distinction between two qualitatively different physical phases. Physical phases are regions in the space of parameters characterizing a theory (such as, for instance, particle masses, coupling constants, or temperature) in the interior of which the expectation values of macroscopic observables (derivatives of the Gibbs potential), written as functions of the parameters, vary only analytically. Boundaries between the different phases are called \textit{phase transitions}.\footnote{Alternatively, one may reserve the notion of a phase transition for the physical \textit{process} of crossing a phase boundary. This is the sense in which, for instance, cosmologists speak of phase transitions in the early universe. For a detailed and rigorous account of phase transitions in the sense of phase boundaries, see \citep{sewell}, Chapter 4. Here I gloss over the difficulties of giving a rigorous account of thermodynamic notions such as the Gibbs potential in the relativistic, quantum field theoretical, context, assuming that at least for all practical purposes these difficulties can be met.} Formulated in terms of this notion, the criterion for gauge symmetry breaking to qualify as a ``natural phenomenon'' stated above translates into the statement that it does so just in case the transition between broken and unbroken gauge symmetry coincides with a phase transition. Cases of SSB on the level of observables automatically count as natural phenomena in this sense, at least if there is a symmetry breaking order parameter in form of the expectation value of a macroscopic observable, which seems to be the case in all the typical cases of practical interest. In view of this tight connection between phase transitions and symmetry breaking it is not so surprising that the vocabulary of SSB is of crucial importance for our understanding and classification of phase transitions. An example of a phase transition that is accompanied by a change of a symmetry from broken to unbroken is the transition between a ferromagnetic and a paramagnetic phase of a magnetic material where the total magnetic moment of the system serves as an order parameter. This quantity is zero throughout the unbroken (``symmetric'') regime but becomes nonzero in the broken regime and therefore must exhibit a non-analyticity (that is, a cusp or a jump) where the symmetry breaking occurs. For the breaking of a gauge symmetry, in contrast, it is not immediately clear on conceptual grounds whether it is necessarily accompanied by a non-analyticity on the level of observables, that is, by a phase transition. A more detailed investigation is required to decide whether specific instances of broken gauge symmetries can count as natural phenomena in that sense. 

For the case of Bose-Einstein condensation discussed in the previous section this question is settled rather easily. We saw that the ground states $\Omega_\theta$ break the gauge symmetry $\alpha^\lambda$ for a free Bose gas at zero temperature. For temperatures $T$ substantially higher than $T=0$, however, the situation looks entirely different. Above a certain \textit{critical temperature} $T_c$ one finds that the expectation value $\Omega(\psi)$ vanishes, which may serve as a symmetry breaking order parameter, signalling that the gauge symmetry is unbroken above $T_c$. The most interesting question for present purposes is whether observable properties of the free Bose gas above the critical temperature $T_c$ are qualitatively different from those below $T_c$. Clearly they are: Thermodynamic quantities such as the compressibility of the gas (which is infinite below $T_c$ in the non-interacting case and nonzero yet finite above $T_c$) show qualitative differences below and above $T_c$, and the temperature dependence of its specific heat exhibits a cusp at $T_c$. Since for a free (i.\ e. non-interacting) system the many-particle states are just symmetrized products of the single-particle states, the microscopic origin of these features can be analysed in terms of occupation numbers of the single-particle states of the free bosons. For temperatures $T<T_c$ below the critical temperature the occupation number $n_0$ of the single-particle ground state (that is, the ground state for a single boson in the same volume) diverges, so that the ratio $n_0/N$ remains finite even when the total number of particles $N$ goes to infinity. At zero temperature, all particles have ``condensed'' into the single-particle ground state, so that $n_0=N=n\cdot V$, where $n$ is the particle density introduced in Eq.\ (\ref{Omega}). For temperatures above the critical temperature $T_c$, in contrast, $n_0/N$ goes to zero as $N$ approaches infinity. The ``condensation'' of particles into the single-particle ground state vanishes together with the restoration of global gauge symmetry, as becomes manifest in the fact that $n_0/N$ can be expressed in terms of the symmetry breaking order parameter. Therefore, in the case of Bose-Einstein condensation of a free Bose gas the distinction between broken and unbroken gauge symmetry corresponds exactly to a distinction on the level of macroscopic observables insofar as situations which are described in terms of broken gauge symmetry are separated by a phase transition from situations described in terms of unbroken gauge symmetry. In Section 6 of this paper I shall argue that this does not always hold for instances of symmetry breaking in gauge theories so that these do not (in general) qualify as natural phenomena in the (weak) sense introduced before in terms of phase transitions.

\section{Local gauge symmetry breaking---the classical perspective}  

In this section, I briefly review the textbook account of the Higgs mechanism in classical terms as a spontaneously broken local gauge symmetry. To see the underlying idea, it suffices to consider, as an example, the Lagrangean of the Abelian Higgs model defined by  
\begin{eqnarray}  
\mathcal L=  D_\mu\phi^*D^\mu\phi-V(\phi)-\frac{1}{4}F_{\mu\nu}F^{\mu\nu}\,,  
\label{HiggsLagrangean}  
\end{eqnarray}  
which exhibits a local $U(1)$ gauge symmetry in that it is invariant under gauge transformations of the form  
\begin{eqnarray}\label{trafos}  
\phi(x)\mapsto e^{i\alpha(x)}\phi(x), \hspace{1cm} A_\mu(x)\mapsto A_\mu(x)-\frac{1}{e}\partial_\mu\alpha(x)\,.  
\end{eqnarray}  
The covariant derivative $D_\mu$ is defined as $D_\mu=\partial_\mu+ieA_\mu$, and the potential $V(\phi)$ in Eq. (\ref{HiggsLagrangean}) is given by  
\begin{eqnarray}  
V(\phi)=m_0^2 \phi^*\phi + \lambda_0 (\phi^*\phi)^2\,,  
\end{eqnarray}  
where the quartic coupling is assumed to be positive, $\lambda_0>0$. If the coefficient of the term quadratic in the fields is negative, that is, if $m_0^2<0$, the potential $V$ has a minimum at a nonzero value of the Higgs field $\phi$, namely, $|\phi|^2=-\frac{m_0^2}{2\lambda_0}$ .  

In this case, the classical ground states of the theory are configurations of the fields $\phi$ and $A_\mu$ of the form  
\begin{eqnarray}\label{configs}  
\phi(x)=e^{i\theta(x)}v/\sqrt{2}, \hspace{1cm} A_\mu(x)=-\frac{1}{e}\partial_\mu\theta(x) ,  
\end{eqnarray}  
where $\theta(x)$ is an arbitrary real-valued function of space and time and $v=\sqrt{-\frac{m_0^2}{\lambda_0}}$. For any two field configurations of the form Eq.\ (\ref{configs}) there exist gauge transformations of the form Eq.\ (\ref{trafos}) that transform them into one another, so all these configurations are physically equivalent. However, since $v\neq0$, the transformations (\ref{trafos}) do not act trivially on these configurations, that is, none of the field configurations (\ref{configs}) is itself invariant under local gauge transformations. This means that local gauge symmetry is indeed spontaneously broken in any classical ground state of (\ref{HiggsLagrangean}).  

In order to extract the physical content of the theory defined by the Lagrangean (\ref{HiggsLagrangean}), it is useful to perform the field redefinition
\begin{eqnarray}\label{elimtrafo}  
\phi(x)=e^{i\theta(x)}\rho(x)&\mapsto&\rho(x)\nonumber\,,\\  
A_\mu(x)&\mapsto& A_\mu(x)+\frac{1}{e}\partial_\mu\theta(x)\equiv B_\mu(x)\,,  
\end{eqnarray}  
which makes it possible to eliminate the $\theta$-field from the Lagrangean, which thereby becomes  
\begin{eqnarray}  
\mathcal L=  \partial_\mu\rho\partial^\mu\rho-V(\rho)+e^2\rho^2B_\mu B^\mu-\frac{1}{4}B_{\mu\nu}B^{\mu\nu}\,, 
\label{HiggsLagrangeanrev}  
\end{eqnarray}
where $B_{\mu\nu}\equiv\partial_\mu B_\nu-\partial_\nu B_\mu$ has been defined.

Expanding the field $\rho$ around its expectation value as $\rho=(v+\eta)/\sqrt{2}$ and neglecting terms which are of third or higher order in the fields $\eta$ and $B_\mu$ one obtains  
\begin{eqnarray}  
\mathcal L^{(2)}=  \frac{1}{2}\left(\partial_\mu\eta\partial^\mu\eta+2m_0^2\eta^2\right)+\frac{1}{2}e^2v^2B_\mu B^\mu-\frac{1}{4}B_{\mu\nu}B^{\mu\nu}\,.  
\label{HiggsLagrangeanfinal}  
\end{eqnarray}  
The characteristic physical properties of the theory defined by this Lagrangean can easily be read off in that it describes a massive vector boson $B_\mu$ with a mass $M_B=ev$ and a massive scalar boson $\eta$ with mass $\sqrt{-2m_0^2}$. The real field $\theta$, which would have played the role of a Goldstone boson in the case of an invariance under \textit{global} gauge symmetries, has been eliminated, and this shows that there are no massless scalar particles in the theory. According to how this is often expressed, the Goldstone boson has been ``eaten'' by the gauge field. The Lagrangean (\ref{HiggsLagrangeanfinal}) contains only gauge invariant fields\footnote{Equivalently, one could have arrived at a Lagrangean of exactly the same form by imposing the unitary gauge $\theta=0$.}, and, from a classical point of view at least, one could have defined the theory directly in terms of these without introducing gauge symmetry at all.\footnote{See \citep{struyve} for a detailed discussion of these questions. If one chooses to use only gauge invariant fields, however, one has to pay careful attention to the constraints for the variable $\eta$ occurring in Eq.\ (\ref{HiggsLagrangeanfinal}), see \citep{struyve}, Section 4, and \citep{strocchi}, p.\ 194. The analysis in terms of the reduced phase space approach given in \citep{struyve}, Section 7, avoids this problem.} Classically, as we see, the Higgs mechanism can be spelled out either in terms of broken local gauge symmetry or without introducing gauge symmetry in the first place. In the formulation using local gauge symmetry, as discussed before, it is broken in any classical ground state.  

In the electroweak theory part of the standard model the implementation of the Higgs mechanism is slightly more complicated than in the case just discussed in that the broken local symmetry is a (non-Abelian) $SU(2)\times U(1)$ symmetry instead of the simpler (Abelian) $U(1)$ symmetry of our example.  Moreover, the $SU(2)\times U(1)$ symmetry is not completely broken by the Higgs field, but only up to a residual $U(1)$ symmetry, which coincides with the gauge symmetry of electromagnetism. Despite these important conceptual differences, however, the conclusion just established that the Higgs mechanism can be described as a case of a spontaneously broken local gauge symmetry is not affected and remains correct for the classical version of the electroweak theory. Leaving aside the classical context from now, I turn to the fate of spontaneously broken local gauge symmetry in \textit{quantized} gauge theories.

\section{Quantization without gauge fixing}
At present we do not have any rigorous formulation of quantum gauge theories in the framework of the algebraic approach to quantum theories, so the status of symmetry breaking in quantized gauge theories has to be discussed within a different framework. Since functional integral quantization seems to be the most common and most convenient approach to the quantization of gauge theories, especially in the non-Abelian case, it is taken as the basis for the discussion of symmetry breaking in quantized gauge theories in the present and following sections. The existence of non-vanishing symmetry breaking order parameter in the sense of Eq.\ (\ref{orderparam}) provides the criterion for SSB in this context.

In the functional integral formulation of quantum field theory, all expectation values of the observables and fields are obtained as derivatives of a generating functional $W[\eta_i]$ that depends on the so-called ``source fields'' $\eta_i$. In the case of a gauge theory with gauge field $A_\mu$ and scalar field $\phi$ this functional can be written as a functional integral (that is, as an integral over all possible field configurations) of the form  
\begin{eqnarray}\label{funcint}  
W[\eta,J]=N\int\mathcal D\phi\mathcal D A_\mu\,\exp\left( i\int {\rm d}^4x(\mathcal {L}+\eta\phi+J_\mu A^\mu) \right)\,,
\end{eqnarray}  
where $N$ is a normalization constant, $\mathcal {L}$ is the Lagrangean of the theory to be quantized and $S=\int {\rm d}^4x\mathcal {L}$ the corresponding action. Correlation functions, which include the expectation values of gauge-dependent quantities that may serve as symmetry breaking order parameters (such as $\langle\phi\rangle$, where $\phi$ is the Higgs field), are obtained from $W[\eta,J]$ by taking derivatives with respect to the source fields $\eta$ and $J$ and then setting them to zero.  

The expression (\ref{funcint}) for $W[\eta,J]$ involves an integral over all possible configurations of the fields $\phi$ and $A_\mu$, which means that each gauge-equivalent class of field configurations is integrated over infinitely often. As a result, the integral in Eq. (\ref{funcint}) diverges in a ``vicious way'' in that the inverse free propagator, a function that is contained in the exponent of the integrand, cannot be inverted to obtain the free propagator itself that is required as a starting point for perturbative calculations.\footnote{The inverse free propagator of the gauge fields can be thought of as the coefficient of the part in the Lagrangean $\mathcal L$ which is quadratic in the gauge fields. For the Abelian case this part of the Lagrangean is given by $-\frac{1}{4}\left(\partial_\mu A_\nu-\partial_\nu A_\mu\right)^2$, and the resulting inverse free propagator for the gauge field is, in momentum representation, given by $\eta_{\mu\nu}k^2-k_\mu k_\nu$ (where $\eta_{\mu\nu}$ corresponds to the matrix ${\rm diag}[1,-1,-1,-1]$). As remarked in the main text, the operator $\eta_{\mu\nu}k^2-k_\mu k_\nu$ is not invertible, which can be seen from the fact that it has $k^\nu$ as an eigenvector with eigenvalue zero.} Non-perturbative calculations that do not require the inverse free propagator in the exponent of the functional integral Eq.\ (\ref{funcint}) to be invertible can be performed by starting from Eq.\ (\ref{funcint}), but this requires the setting of lattice gauge theory, where the gauge theory to be quantized is not formulated on the space-time continuum but rather on a discrete lattice of space-time points.  

There are at least two different types of possible reactions to this problem, which I consider in the present and following sections, respectively. The first is to choose a non-perturbative route and try to determine physical features of the quantized gauge theory defined by the functional integral in ways that do not require a free propagator. For many practical purposes, the most convenient such approach is lattice quantization, where one considers space-time not as a continuum but as a lattice of discrete space-time points and extrapolates the results obtained to the continuum case by letting the lattice spacing go to zero. The second reaction, discussed in the following section, is gauge fixing---the insertion of terms in the functional integral that violate gauge invariance, but in such a way that correlation functions for gauge invariant quantities are independent of the choice of gauge fixing terms. Since local gauge symmetry is explicitly broken by gauge fixing terms, one has to consider the option without gauge fixing to assess the fate of local gauge symmetry breaking in quantized gauge theories. The next section will focus on the breaking of post-gauge fixing remnant \textit{global} gauge symmetries that, depending on the choice of gauge, survive in the presence of gauge fixing terms. 

There are different approaches to determine the properties of quantized gauge theories without gauge fixing, but the most common framework for carrying out such computations is lattice gauge theory\footnote{For an alternative approach that does not use gauge fixing see the version of continuum perturbation theory used in \citep{buchmueller} to study the electroweak phase transition in gauge invariant terms.}, where scalar and fermion fields are defined on a lattice representing discretized space-time, and the gauge fields are defined on the links connecting the lattice sites.\footnote{For the earliest introduction of lattice gauge models, see \citep{wegner}. Lattice gauge theory as sketched in this paragraph was essentially invented by Wilson, see \citep{wilson}. For a gentle modern introduction to lattice gauge theory, see \citep{muenster}.} By considering finite lattices with periodic boundary conditions the functional integrals can be evaluated explicitly in a non-perturbative way, so that no expansion of expressions like Eq.\ (\ref{funcint}) needs to be made that requires a free propagator for the gauge field. One important virtue of this approach is that it allows numerical computations that do not require the couplings constants describing the gauge interactions to be small (as perturbative calculations usually do), another is that it provides a framework to derive rigorous results. This sets it off from the continuum formulation, where functional integrals such as Eq.\ (\ref{funcint}) are used in a mere formal manner without any rigorous mathematical underpinning. In the absence of gauge fixing terms, local gauge symmetry persists after quantization, and it is possible to discuss whether it can be spontaneously broken.

At least according to current state-of-the-art knowledge of quantized gauge theories the answer to this question is negative: Local gauge symmetry cannot be spontaneously broken at all in a gauge quantum field theory in that the expectation values of arbitrary gauge-dependent combinations of fields are always zero, in particular the expectation value of any Higgs-type scalar field of the theory harbours such a field. In the framework of lattice gauge theory this result can be derived in a rigorous fashion and, as such, is widely known as \textit{Elitzur's theorem}.\footnote{Elitzur proved the theorem for the case of a Higgs field with fixed modulus, see \citep{elitzur}. The result was generalized to the case of a Higgs field with variable modulus by de Angelis, de Falco and Guerra, see \citep{DDG}. See \citep{itzykson}, Chapter 6.1.3, for a useful textbook version of the proof.}

As already remarked, in the continuum case the functional integral formulation is not mathematically well-defined, so there exists no rigorous derivation of an impossibility result for spontaneously breaking local gauge symmetries in this context. There are no indications, however, that the impossibility of spontaneously breaking local gauge symmetries might cease to hold in future continuum formulations of quantized gauge theories one may speculate about. The crucial feature of local gauge symmetries used in the proof of Elitzur's theorem is that for any finite volume of space-time local gauge transformations can always be chosen such as to act non-trivially only in that finite volume (and to reduce to the identity transformation everywhere else). In virtue of this feature of local gauge symmetries it is impossible to impose a symmetry breaking order parameter for the local gauge symmetry by implementing symmetry breaking boundary conditions on a finite volume boundary and subsequently removing the volume boundary to space-time-like infinity.\footnote{See \citep{strocchiold}, Chapter II 2.5, and \citep{frohlich}, Section 3, for more on the importance of this feature of local symmetries in the derivation of Elitzur's theorem.} Evidently, this feature holds independently of whether the theory is formulated on a lattice or in the continuum. Any non-trivial \textit{global} symmetry transformation, in contrast, acts non-trivally on the degrees of freedom of \textit{all} of space-time, not only in a finite space-time volume. This blocks the derivation of an analogous result for the case of a global symmetry, so the impossibility of spontaneously breaking local gauge symmetries according to Elitzur's theorem is not so much a consequence of the general unobservability of gauge transformations but has to do with the specific features of \textit{local} symmetries. To sum up, while there can be no \textit{mathematical} guarantee that local gauge symmetry might not be spontaneously broken in some future continuum formulation of gauge quantum field theory (even though this seems unlikely), the notion of a spontaneously broken local gauge symmetry is not supported by the leading frameworks of gauge theory quantization which we presently have.

Recognizing this, one may wonder whether the Higgs mechanism may perhaps not work as an account of mass generation in the standard model since textbook expositions of the Higgs mechanism are commonly based on the notion of a spontaneously broken local gauge symmetry. Fortunately, however, as demonstrated by Fr\"ohlich, Morchio, and Strocchi\footnote{See \citep{frohlich}.}, such fears are ungrounded, since the physical phenomena which are usually associated with the Higgs mechanism can be recovered in terms of an approach that uses only entirely gauge invariant fields. They develop a recipe for constructing gauge invariant combinations of the Higgs and gauge fields that allows to reformulate any gauge theory in terms of such gauge invariant combinations. Observable quantities, such as the (Yukawa) couplings between the gauge bosons and fermions in the conventional formulation, are obtained as functions of expectation values of these gauge invariant fields. In particular, Fr\"ohlich et al. provide a list of gauge invariant quantities that are non-vanishing and correspond directly to quantities identifiable with the particle masses in the conventional formulation using gauge dependent fields. So, mass generation through the Higgs mechanism can get along very well without assuming a nonzero expectation value for any gauge dependent combination of fields, in particular not for the Higgs field itself.

One may conclude from the fact that mass generation through the Higgs mechanism, as demonstrated by Fr\"ohlich, Morchio and Strocchi, can be accounted for in terms of gauge invariant fields that to characterize the Higgs mechanism as a spontaneously broken local gauge symmetry is, as Smeenk puts it, a ``relatively benign case of abuse''\footnote{See \citep{smeenk}, p.\ 498.} of terminology. An alternative conclusion to draw, however, would be that the abuse of terminology involved in characterizing the Higgs mechanism as a spontaneously broken local gauge symmetry is not so benign---after all, the notion is vacuous according to our currently best conceptual frameworks--- but that despite its being vacuous the notion of a spontaneously broken local gauge symmetry has an important heuristic value, at least historically, and may still be useful for semi-classical calculations.  

Another worry that might be brought up by Elitzur's theorem is that if we do not dispose of the notion of a spontaneously broken local gauge symmetry, we can no longer make the important distinction between cases where local gauge symmetry is broken and cases where it is unbroken (``restored''). This distinction, however, is apparently crucial to describe the \textit{electroweak phase transition}, a phase transition between two different phases of the universe as described by the electroweak theory at different values of the fundamental parameters such as temperature and the Higgs boson mass. This transition is widely believed to have actually taken place as temperature decreased in the history of the early universe so that it supposedly evolved from a phase where the $SU(2)\times U(1)$ local gauge symmetry of electroweak theory is unbroken to the phase in which we now exist, where that symmetry is allegedly broken.\footnote{Detailed calculations (see, for instance, \citep{kajantie}) have shown that for values of the Higgs mass not excluded by experiment the electroweak phase transition is actually not a real phase transition (in the sense of an abrupt change in thermodynamic quantities) but rather a steep crossover between two qualitatively different regimes of electroweak theory, meaning that at least some expectation values of observables vary very strongly (yet analytically) from one regime to the other. In the context of the present paper, however, the question of whether, for realistic values of the Higgs mass, the electroweak phase transition is a genuine phase transition or rather a continuous crossover is not important since we are concerned here with the more general question of whether the notion of a spontaneously broken local gauge symmetry is needed to give meaning to the distinction between the high and low temperature phases of the electroweak theory, which are sharply separated for \textit{some} values of the Higgs mass.} In the phase where the electroweak symmetry is said to be unbroken (``restored'') the electron and the neutrino are not yet distinguishable in that they correspond to degenerate states of one and the same particle. At the present state of the universe, in contrast, there is obviously a substantive physical difference between the electron and the neutrino, so the supposed phase transition seems to have taken the universe from one phase to another, qualitatively very different, one. Do we have to conclude from Elitzur's theorem that the very idea of an electroweak phase transition rests on an error in that there cannot be a transition from a situation where electroweak symmetry is unbroken to a situation where it is broken since electroweak symmetry can never be broken at all?  

Fortunately, this conclusion need not be drawn since the electroweak phase transition, just as the Higgs mechanism itself, can be described in purely gauge invariant terms. An example of an observable, that is, gauge invariant quantity that may quite drastically change at the phase transition is the expectation value $\langle\phi^*\phi\rangle$ (where $\phi$ is the Higgs field), which, if displayed as a function of parameters such as temperature and the Higgs boson mass, exhibits a ``jump'' across the planes in the phase diagram where the electroweak phase transition occurs.\footnote{See, for instance, \citep{buchmueller}, pp.\ 134-6.} From the fact that phase transitions are often accompanied by the breaking (or restoration) of certain symmetries and the fact that the electroweak phase transition is often associated with ``electroweak symmetry breaking'' one might mistakenly conclude that there is an incompatibility between the impossibility of spontaneously breaking local gauge symmetries and the electroweak phase transition. As we have just seen, however, this is not the case, for the distinction between the two different phases, one where electroweak symmetry is allegedly broken and one where it is allegedly unbroken, can be made in an entirely gauge invariant way so that the dubious notion of a spontaneously broken local gauge symmetry is altogether avoided. Phase transitions are indeed often accompanied by instances of symmetry breaking, but the definition of a phase transition in terms of non-analytic behaviour of observable quantities does not require symmetry breaking. The electroweak phase transition, as we see, is a case in point.  

The topic of phase transitions in gauge theories will concern us again in the following section while discussing the role of spontaneous symmetry breaking in the presence of gauge fixing terms.

\section{Gauge fixing and symmetry breaking}  
Having discussed the quantization of gauge theories without gauge fixing in the lattice formulation of gauge theories, I now turn to their quantization by means of gauge fixing terms, which makes it possible to perform perturbative computations using the diagrammatic techniques invented by Feynman in the continuum as well as on the lattice. In classical gauge theories, gauge fixing amounts to the implementation of constraints for the Higgs and gauge fields such as, for instance, the unitary gauge mentioned in Section 4, which fixes the phase of the Higgs field at a constant value, say zero, at any space-time point. For the Higgs field in the Abelian Higgs model discussed in Section 4, which can be written as $\phi(x)=e^{i\theta(x)}\rho(x)$, this means setting $\theta(x)=0$ for all $x$. Other choices of gauge fixings that tend to be better suited for practical calculations include the Coulomb gauge, defined by $\partial_i A^i=0$ (where the summation is over spatial indices only), and the Lorenz gauge, defined by $\partial_\mu A^\mu=0$.  

In the functional integral formulation of quantum field theory, gauge fixing is implemented in the form of field-valued Dirac-$\delta$-functions in the functional integral. The introduction of these $\delta$-functions can be seen as part of a change of integration variables involving a Jacobi determinant, the so-called Faddeev-Popov determinant $\Delta(A)$, and it requires, at least in certain gauges, the introduction of additional, purely formal, fields as integrations variables. These are the so-called ghost fields, which do not correspond to any physical degrees of freedom.\footnote{This can be seen, for instance, from the fact that ghost fields formally correspond to spinless fermion fields the physical existence of which is excluded by the spin-statistics theorem.} The original gauge invariant action $S$ of the gauge theory to be quantized (corresponding to the integral in the exponent of Eq.\ \ref{funcint}) is replaced by an ``effective'' action $S_{\mathit{eff}}$ of the form  
\begin{eqnarray}\label{Sgf} 
S_{\mathit{eff}}=S+S_{\mathit{gf}}+S_{\mathit{ghost}}\,,  
\end{eqnarray}  
where $S_{gf}$ implements the gauge fixing in that it contains the gauge fixing constraint and $S_{ghost}$ is an additional term in the presence of ghost fields.  

The gauge fixing term $S_{\mathit{gf}}$ in the ``effective'' action $S_{\mathit{eff}}$ explicitly violates local gauge invariance in that some non-gauge invariant term is inserted in the functional integral. This is done in such a way that the physical content of the theory remains unchanged, so the gauge fixing does not have any physical significance whatsoever. However, the way in which local gauge invariance is violated by this procedure depends on the choice of gauge fixing made. One possibility is that the gauge freedom is \textit{completely} eliminated by the gauge fixing in the sense that out of any class of gauge-equivalent field configurations exactly one is singled out by the gauge fixing constraint. This is the case for the unitary gauge, which, in the case of the locally $U(1)$-symmetric Abelian Higgs model discussed before, is given by $\theta(x)=0$. Here, local gauge symmetry is eliminated completely (and explicitly) at the level of the ``effective'' action $S_{\mathit{eff}}$, so \textit{spontaneous} symmetry breaking cannot occur any more, for there simply is no unbroken symmetry left to be broken.

For other choices of gauge fixing terms, however, the action $S_{\mathit{eff}}$ can still be invariant under symmetry transformations corresponding to some finite-parameter subgroup of the original infinite-parameter local gauge group. In the presence of gauge fixing terms of this class, the action $S_{\mathit{eff}}$ still exhibits certain \textit{global} gauge symmetries, but no longer a local one. The spontaneous breaking of global symmetries is not forbidden by Elitzur's theorem, and indeed the breaking of these remnant global gauge symmetries is a common phenomenon in gauge theories in the presence of gauge fixing. In what follows, I will refer to it as the spontaneous breaking of ``global subgroups'' of the original, local, gauge group or just ``remnant symmetry breaking''. It can also be studied in the formulation without gauge fixing, discussed in the previous section, by introducing fields which depend on the spacetime variable $x$ not only in an explicit manner, but also implicitly, via an additional dependence on the gauge fields. An example of such a field is\footnote{The example taken is Eq.\ (1.1) in \citep{caudy}.}  
\begin{eqnarray}\label{Phi}  
\Phi(x;A)=g(x;A)\phi(x)\,,  
\end{eqnarray}  
where $\phi(x)$ is the Higgs field and $g(x;A)$ is a gauge transformation, which takes the gauge field into a chosen gauge such as the Coulomb or Landau gauge. The so defined $\Phi(x;A)$ has a nonzero expectation value just in case the Higgs field $\phi(x)$ itself has a nonzero expectation value for the respective choice of gauge fixing, that is, for the choices mentioned, in the Coulomb or Landau gauge. It thus functions as a symmetry breaking order parameter for the remnant global gauge symmetry defined by the transformation $g(x;A)$.

Since we do not presently have any notion of a spontaneously broken local gauge symmetry in a gauge quantum field theory, the breaking of these remnant global subgroups is the only sense of gauge symmetry breaking that remains to be elucidated.\footnote{There are other symmetries besides local gauge symmetries and their global subgroups which can be broken in quantized gauge theories such as, for instance, chiral symmetry in QCD or centre symmetry in non-Abelian gauge theories (the centre of a group is the set of elements which commutes with all other elements), which seems to be linked to the confinement-deconfinement phase transition, see \citep{greensite}. The present paper is not concerned with the breaking of these symmetries but only with that of local gauge symmetries and their global subgroups.} To answer the question of whether gauge symmetry breaking in quantized gauge theories can count as a natural phenomenon in the sense spelled out in Section 3 in terms of phase transitions, we therefore have to investigate whether the distinction between broken and unbroken remnant gauge symmetry always lines up with a contrast between distinct physical phases. We have to ask, in other words, whether the transition from unbroken to broken global subgroups is always accompanied by an abrupt change in the expectation values of some observables. 

Even though there does not seem to be any rigorous statement about the relation between remnant symmetry breaking and the occurrence of phase transitions in gauge theories, there is strong evidence, based on a combination of exact and numerical results, that there is \textit{no} rigid connection between the two and that, therefore, remnant gauge symmetry breaking does not in general qualify as a natural phenomenon in the sense specified in Section 3. A particularly illuminating discussion of the relation between the breaking of remnant subgroups and phase transitions is given by Caudy and Greensite in the context of a study of an $SU(2)$-symmetric lattice gauge model with a fixed-modulus Higgs field.\footnote{See \citep{caudy}. More precisely, their results are for a model with a fixed-modulus Higgs field in the fundamental colour representation. Their results clearly show that in a certain range of parameters the system exhibits the typical features of a ``Higgs phase'' such as, for instance, the appearance of a massive spectrum associated with the gauge field degrees of freedom, even though there is no ``Mexican hat potential'' (which makes sense only for a Higgs field with a variable modulus).} For this model, there is robust numerical evidence that there exists, in a limited region of the phase diagram, a phase transition between a ``Higgs phase'', where the spectrum exhibits a gauge boson mass, and a ``non-Higgs phase'', where there is no such mass and the properties of the model are more similar to those of quantum chromodynamics (QCD) in the presence of confinement.\footnote{See \citep{greensite} for an introduction to the problem of confinement that includes an in-depth discussion of how confinement should be defined in the first place.} The main conclusion drawn by Caudy and Greensite from their results is that there is no \textit{general} agreement between the two transition lines (that between the different phases and that between broken and unbroken remnant gauge symmetry), even though for \textit{some} values of the parameters of the model the transition between the two phases does coincide with that between a regime where remnant symmetry is broken and one where it is unbroken.

This conclusion has two distinct interesting aspects the first of which is that, according to the results reported by Caudy and Greensite, in both the
Coulomb and the Landau gauges, part of the separation line between broken and unbroken gauge symmetry occurs at parameters where, on grounds of an earlier exact result due to Fradkin and Shenker \citeyearpar{fradkin}, the existence of an accompanying phase transition can be determinately excluded. According to Fradkin and Shenker, there exists a continuous path in parameter space connecting the regime which displays features typical of the Higgs mechanism and the regime which displays features typical of confinement, along which the expectation values of all observables vary analytically.\footnote{An interesting line can be drawn from the Fradkin-Shenker result to more recent developments involving duality in supersymmetric gauge theories and their ramifications for string theory. See \citep{seiberg} for an introduction to supersymmetric gauge theories that makes this connection.} This implies that any phase boundary which partly separates the two regimes must have an endpoint in parameter space beyond which all expectation values vary only analytically, analogously to the case of the phase boundary between the liquid and gaseous phases in a typical phase diagram of ordinary matter, where beyond the critical endpoint of the phase boundary between the liquid and gaseous phases the distinction ceases to be sharp and becomes gradual. According to the results obtained by Caudy and Greensite, the distinction between regimes with broken and unbroken remnant gauge symmetries coincides \textit{partly} with the phase transition between the Higgs and confinement regimes, but it continues beyond the endpoint of the phase transition line into a region of parameter space where no phase transitions occur and all observables vary smoothly.

Remnant gauge symmetry breaking, to conclude, is not in general linked to a transition between distinct physical phases as in the Bose-Einstein case discussed in Section 3 in that the transition between broken and unbroken remnant subgroups can occur in regimes where all observables vary analytically. This shows that remnant symmetry breaking is not in general a natural phenomenon in the sense specified in Section 3. A second interesting aspect of the conclusions presented by Caudy and Greensite is that, according to their results, the values of parameters of the theory for which there is a transition between unbroken and broken gauge symmetry depend on the \textit{choice} of remnant subgroup, that is, if gauge fixing is used, on the choice of gauge fixing terms. They summarize this observation in the statement that gauge symmetry breaking in gauge theories is ``ambiguous'' in the sense that whether or not remnant gauge symmetry is broken for a specific choice of parameters in general depends on the (from a physical point of view) arbitrary choice of remnant subgroup. This observation illustrates further why remnant symmetry breaking does not deserve to be called a ``natural phenomenon'' in that whether or not it occurs for a given choice of parameters depends on the unphysical (gauge) freedom of description. 

In the following section I consider some consequences of the considerations presented in this and the previous sections for philosophical debates about the interpretation of gauge symmetries and their breaking.

\section{Philosophical implications}  
The considerations on gauge symmetry breaking presented in the previous sections have interesting philosophical ramifications. In particular, they imply that some interpretive claims about gauge symmetries and their breaking in the literature are misleading. I discuss three examples of such claims.  

The first example is Peter Kosso's contention that broken gauge symmetries belong to the class of cases where ``the relevant laws of nature are exactly symmetric, but the phenomena expressing these laws are not.''\footnote{See \citep{kosso}, p. 359.} That this characterization cannot really be adequate follows already from the fact that gauge symmetries have no physical instantiations. If a theory such as that of the Bose gas discussed in Section 3 has ground states that break (global) gauge symmetry, all these ground states are still physically equivalent in that with respect to observable quantities they all assign the same expectation values. Kosso's question of why we should think that the fundamental interactions of nature are ``gauge symmetric'' even though the phenomena which we observe are not is misleading since there is no asymmetry in the phenomena that is not found in the basic laws due to the fact that gauge symmetries are purely formal and hence unobservable. The defence of the Higgs mechanism as an account of mass generation in the standard model may still raise interesting epistemological challenges, but this has nothing to do with the issue of conjecturing the fundamental laws to be symmetric in a way in which the phenomena we observe are not.  
 
A number of claims on the nature and role of gauge symmetry breaking in gauge theories are based on failure to take into account Elitzur's theorem and the fact that whether or not the Higgs field has a nonzero expectation values depends on the choice of gauge fixing. Margaret Morrison, for instance, argues that the Higgs mechanism is ``based on the idea that even the vacuum state can fail to exhibit the full symmetries of the laws of physics.''\footnote{See \citep{morrison}, p. 356.} As a claim about ideas that have historically played a role in the development of the Higgs mechanism this statement may be true, but Morrison argues further that even from a methodological point of view ``one needs the underlying vacuum assumptions regarding the plenum and degeneracy as part of the `physical' picture.''\footnote{See \citep{morrison}, p. 357.} An integral part of this picture, as she claims, is the thought that here ``we are dealing with fields whose \textit{average value} is non-zero, where the vacuum is said to have a non-zero expectation value.''\footnote{See \citep{morrison}, p. 359.} This statement, as we have seen, is not correct in that, according to our best accounts, the vacuum expectation of the Higgs field \textit{is} actually zero in the absence of any gauge fixing in the quantum case, whereas in the presence of gauge fixing it depends on the choice of gauge fixing which, practical considerations aside, is arbitrary from a physical point of view. Morrison's central conclusion that ``it would be folly to accept a robust physical interpretation of the SSB story''\footnote{See \citep{morrison}, p. 361.} in the electroweak theory is quite plausible (depending on what exactly is meant by ``robust physical interpretation''), but the reason she gives for drawing the conclusion, namely, ``that the various vacuum hypotheses which provide the necessary theoretical foundations are essentially problematic, for both physical and philosophical reasons''\footnote{See loc. cit.}, is not completely convincing. The problematic aspect of the notion of spontaneous symmetry breaking in the context of the $SU(2)\times U(1)$ symmetry of the electroweak theory is not that it is based on a questionable ``theoretical story about the nature of the vacuum''\footnote{See loc. cit.}, but that the $SU(2)\times U(1)$ \textit{local} gauge symmetry is unbroken in the quantum case (according to what we now know of gauge quantization at least), whereas the breaking of remnant subgroups depends on the gauge fixing.

Misunderstandings about the nature and significance of SSB in gauge theories can be found not only among philosophers but also among eminent physicists. Steven Weinberg, for instance, argues in a ground-breaking paper on phase transitions in gauge theories that these phase transitions have the ``philosophical implication'' as regards the ``reality'' of gauge symmetries that ``if a gauge symmetry becomes unbroken for sufficiently high temperature, it becomes difficult to doubt its reality.''\footnote{See \citep{weinberg}, p.\ 3359.} Weinberg's reasoning here seems to be that if gauge symmetries exist in both broken and unbroken forms in such a way that there is a substantial physical difference between the two cases (that is, a phase transition that separates them), these symmetries are the bearers of non-trivial physical properties and, therefore, must be real. Although there may be disagreement about the sense in which gauge symmetries are supposedly established as ``real'' according to this line of thought, it seems clear from the considerations presented in the previous sections that Weinberg's argument fails, whatever exactly it is supposed to show, for several reasons. As we have seen, there is no reason to suppose that \textit{local} gauge symmetry is ever broken in a quantized gauge theory, so one should not expect phase transitions such as the electroweak phase transition to be described in terms of its breaking and the existence of phase transitions cannot have any implications whatsoever for the reality of local gauge symmetries. Remnant global subgroups of local gauge groups, on the other hand, may break spontaneously, but their breaking is ambiguous in that it depends on the gauge and is not necessarily accompanied by a qualitative change in physical properties. It seems therefore problematic to regard these global symmetries as the true bearers of physical properties and thus as ``real'' in a more substantial sense than the original, local, symmetries. The standard view of gauge symmetries as purely formal symmetries which do not have physical instantiations, in particular, is not in the least called into question by the result that there are phase transitions in gauge theories at high temperatures which for certain choices of gauge fixing are accompanied by restorations of remnant global symmetries.

\section{Conclusion}
The aim of this paper has been to clarify the status and significance of gauge symmetry breaking in gauge theories. The classical ground state of the Abelian Higgs model was presented as an example of a state which, if described in terms of gauge variables, exemplifies the spontaneous breaking of local gauge symmetry. Current wisdom of \textit{quantized} gauge theories, in contrast, does not support any notion of a spontaneously broken local gauge symmetry. In the framework of lattice gauge theory, the statement that local gauge symmetries cannot be spontaneously broken can be made rigorous and, as such, is referred to as ``Elitzur's theorem''. Remnant gauge symmetries were introduced as global symmetries with respect to which the action of a gauge theory may remain invariant after gauge fixing. In contrast to local gauge symmetries, remnant global gauge symmetries may break spontaneously. The physical significance of these instances of symmetry breaking was considered by investigating their relation to transitions between distinct physical phases. Based on the results of \citep{caudy}, it was argued that there seems to be no general fixed connection between remnant gauge symmetry breaking and phase transitions in that, first, a transition between broken and unbroken remnant gauge symmetry can exist without any accompanying discontinuous change in the expectation values of observables and, second, the breaking of remnant gauge symmetry may depend on the choice of gauge fixing made.\footnote{Or, equivalently, on the choice of gauge transformation $g(x;A)$ as in Eq.\ (\ref{Phi}), used to define a remnant subgroup of the original local gauge group.}

With respect to the Higgs mechanism the following two conclusions can be drawn from the considerations presented: The first is that the standard textbook characterization of the Higgs mechanism as a spontaneously broken \textit{local} gauge symmetry is misleading according to what we presently know (even though perhaps useful from a heuristic point of view) in that it is valid only for the classical, not for the quantum, case. The second is that while remnant \textit{global} gauge symmetries may indeed be broken in regimes that exhibit the typical features of a ``Higgs phase'', it does not suffice to detect the breaking of a remnant global symmetry to establish that these features actually hold. A direct inspection of gauge invariant quantities remains necessary.

\section*{Acknowledgements}
I would like to thank Kerry McKenzie, Robert Harlander, Dennis Lehmkuhl, Holger Lyre, Michael Kobel, Michael Kr\"amer, Michael St\"oltzner, Ward Struyve and anonymous referees who reviewed this article for many helpful comments. Furthermore, I am grateful to Jeff Greensite, Gernot M\"unster and Franco Strocchi for useful answers to questions I raised.

\end{document}